\documentclass[aps,amsmath,amssymb,prl,floatfix,reprint,superscriptaddress]{revtex4-2}
\usepackage{lipsum} 
\usepackage{bm}
\usepackage{times}
\usepackage{graphicx}
\usepackage{float}
\usepackage{color}
\usepackage{comment}
\usepackage{morefloats}
\usepackage{rotating}

\usepackage{siunitx}
\usepackage{srcltx}
\usepackage{slashed}
\usepackage{subfigure}
\usepackage{multirow}
\usepackage{verbatim}
\usepackage{hyperref}
\usepackage{tabularx}
\usepackage{adjustbox}
\usepackage{comment}
\usepackage[dvipsnames]{xcolor}
\usepackage{nicefrac}
\usepackage{cleveref}

\newcommand*{\dif }{\mathop{}\!\mathrm{d}}
\newcommand{\jpsi}{J/\psi}
\newcommand{\piz}{\pi^{0}}
\newcommand{\ee}{e^{+}e^{-}}
\newcommand{\eexi}{\jpsi\to\Xi^{0}(\to\Lambda\gamma)\bar{\Xi}^0(\to\bar{\Lambda} \piz)}

\newcommand{\SigRadDec}{\Sigma^{+}\to p \gamma}

\newcommand{\XiRadDec} {\Xi^0\to \Lambda\gamma }

\newcommand{\XibHadDec}{\bar{\Xi}^0\to \bar{\Lambda}\piz}
\newcommand{\LDec}{\Lambda\to p\pi^-}
\newcommand{\LbDec}{\bar{\Lambda}\to \bar{p}\pi^+}
\newcommand{\xiz}{\Xi^0}
\newcommand{\xib}{\bar{\Xi}^0}
\newcommand{\jpsiXX}{\jpsi\to\Xi^0\bar{\Xi}^0}

\newcommand{\mrec}{M_{\rm rec}}

\newcommand{\ar}{\alpha_{\gamma}}
\DeclareSIUnit\clight{\text{\ensuremath{c}}}

\hypersetup{
	colorlinks=true,
	allcolors=blue
}

\begin{document}

\title{Status and prospect of weak radiative hyperon decays}
\author{Rui-Xiang Shi}\thanks{These authors contributed to the work equally and should be regarded as co-first authors}
\affiliation{Department of Physics, Guangxi Normal University, Guilin, 541004, China}
\affiliation{Guangxi Key Laboratory of Nuclear Physics and Technology, Guilin, 541004, China}
\author{Zekun Jia}\thanks{These authors contributed to the work equally and should be regarded as co-first authors}
\affiliation{University of Science and Technology of China, Hefei 230026, China}
\affiliation{State Key Laboratory of Particle Detection and Electronics, Hefei, 230026, China}
\author{Li-Sheng Geng}
\affiliation{School of Physics, Beihang University, Beijing 102206, China}
\affiliation{Centrale Pekin, Beihang University, Beijing 100191, China}
\affiliation{Peng Huanwu Collaborative Center for Research and Education, Beihang University, Beijing 100191, China}
\affiliation{Beijing Key Laboratory of Advanced Nuclear Materials and Physics, Beihang University, Beijing 102206, China}
\affiliation{Southern Center for Nuclear-Science Theory (SCNT), Institute of Modern Physics, Chinese Academy of Sciences, Huizhou 516000, China}
\author{Haiping Peng}
\affiliation{University of Science and Technology of China, Hefei 230026, China}
\affiliation{State Key Laboratory of Particle Detection and Electronics, Hefei, 230026, China}
\author{Qiang Zhao}
\affiliation{Institute of High Energy Physics, Chinese Academy of Sciences, Beijing 100049, China}
\affiliation{University of Chinese Academy of Sciences, Beijing 100049, China}
\affiliation{Center for High Energy Physics, Henan Academy of Sciences, Zhengzhou 450046, China}
\author{Xiaorong Zhou}
\affiliation{University of Science and Technology of China, Hefei 230026, China}
\affiliation{State Key Laboratory of Particle Detection and Electronics, Hefei, 230026, China}

\begin{abstract}
Weak radiative hyperon decays represent a rich interplay between weak interactions and the internal structure of baryons, offering profound insights into Quantum Chromodynamics and weak interactions. Recent experimental observations, particularly from BESIII, have revealed deviations from theoretical predictions. These deviations could signal new physics or the need for refined theoretical models incorporating intermediate resonance effects. This review discusses recent theoretical advancements and key experimental findings, focusing on recent measurements from BESIII and their implications for strong interactions and baryon structure.
\end{abstract}


\maketitle


{\it{1.~Introduction.}}  Hyperons are a class of light baryons with one or more strange quarks in their composition. Since their discovery, investigations into their production mechanisms and decay processes have been a hot topic in hadron physics. These investigations are essential because hyperon decays are primarily governed by weak interactions, and several open questions remain regarding the weak decays of hyperons~\cite{Holstein:1999gx,Li:2016tlt}. These unresolved issues are closely connected to the intricate non-perturbative dynamics of strong interactions. Furthermore, weak decays of hyperons provide a unique opportunity to study flavor-changing neutral current (FCNC) processes. Notably, these processes in hyperon weak decays can typically probe energy scales higher than their counterparts in charm or beauty-hadron decays, assuming similar coupling strengths~\cite{Nir:2007xn}. This sensitivity arises due to the presence of the largest Cabibbo factor among these processes. Consequently, weak hyperon decays represent a vital experimental and theoretical venue for exploring the nature of non-perturbative strong interactions and for probing possible new physics beyond the Standard Model (SM).

\begin{figure*}[htpb]
  \centering
   \includegraphics[width=0.24\linewidth]{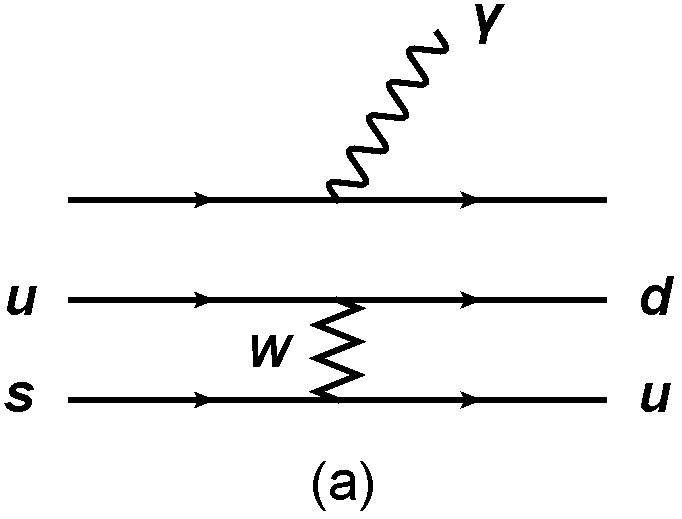}\qquad\qquad
  \includegraphics[width=0.24\linewidth]{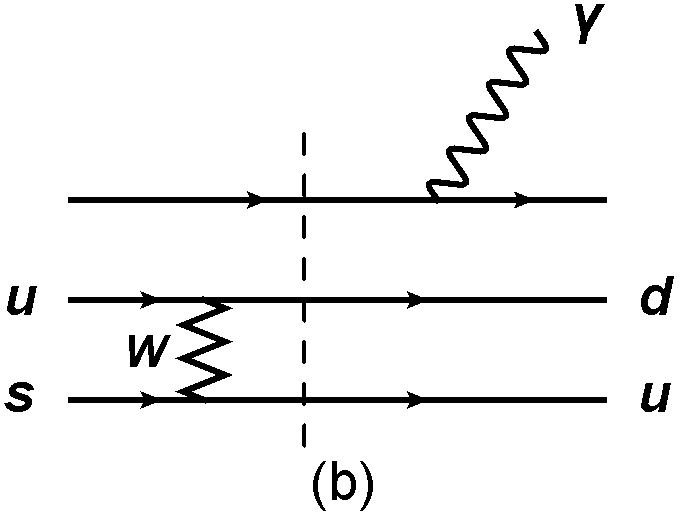}\\
  \includegraphics[width=0.24\linewidth]{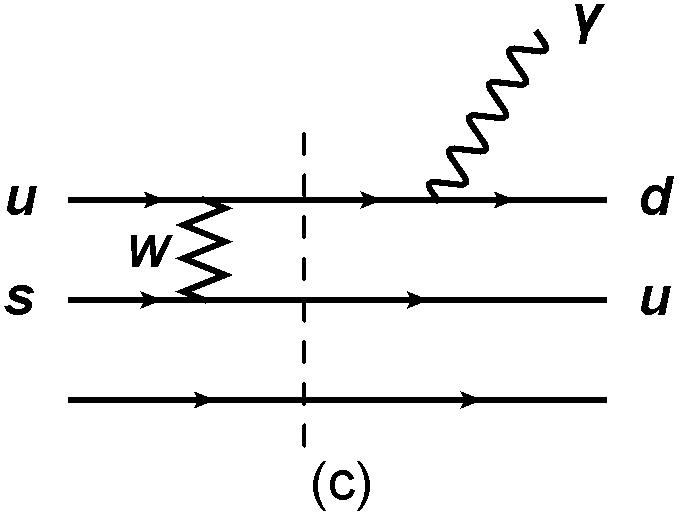}\qquad\qquad
  \includegraphics[width=0.24\linewidth]{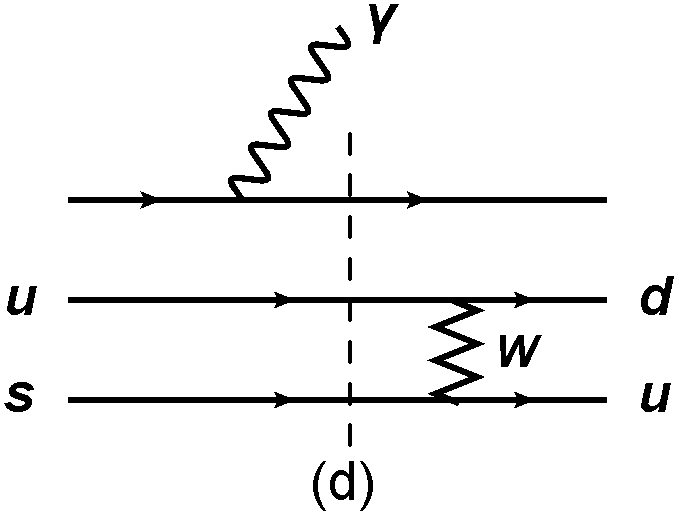}\qquad\qquad
  \includegraphics[width=0.24\linewidth]{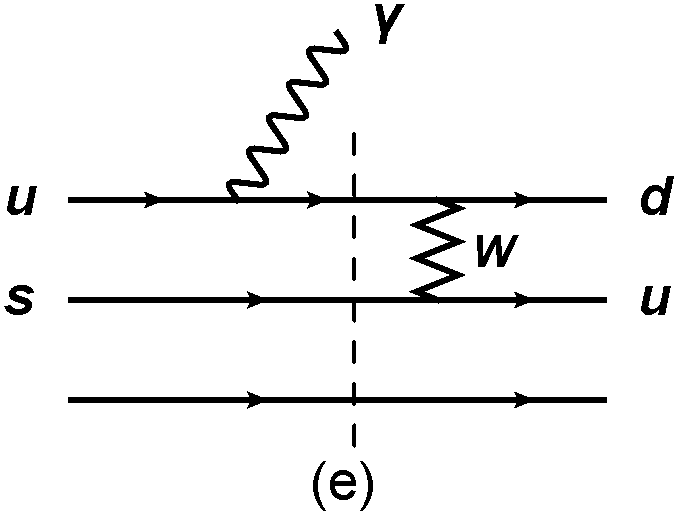}
  \caption{The two-quark transition diagrams for weak radiative hyperon decays. Dashed lines denote that there are intermediate states in these Feynman diagrams.}\label{Fig:QuarkPlots}
\end{figure*}

\begin{figure}[htb]
  \centering
  \includegraphics[width=0.5\linewidth]{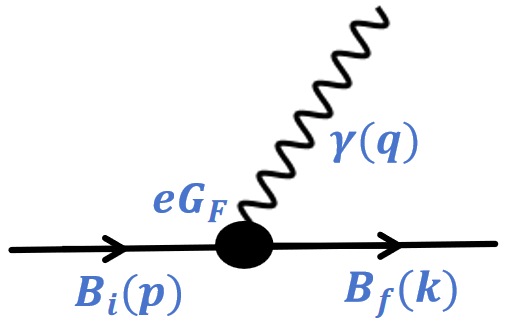}\\
  \caption{The effective diagram for weak radiative hyperon $B_i\to B_f\gamma$ decays and $q=p-k$ in the figure.}\label{Fig:HadronPlot}
\end{figure}

In general, weak decays of hyperons can be categorized into three types: radiative, non-leptonic, and semi-leptonic decays. This review will focus specifically on the weak radiative hyperon decays (WRHDs), which are characterized by the $s\to d(\gamma)$ transition involving only one initial and final quark, or by the $su\to ud(\gamma)$ conversion involving two initial and final quarks~(See Fig.~\ref{Fig:QuarkPlots}). It should be noted that the $s\to d(\gamma)$ transition is via a Penguin loop. Thus, it is highly suppressed in comparison with the tree-level $su\to ud(\gamma)$ conversion. These two-body decays represent a complex interplay of strong, weak, and electromagnetic interactions. Note that the weak radiative decay of the spin-3/2 $\Omega^-$ hyperon is not discussed here due to the scarcity of experimental measurements. Throughout this paper, we will restrict our attention to the $\Delta S=1$ weak radiative decays of the lowest-lying spin-1/2 hyperons. In the real world, there are six distinct decay channels: $\Lambda\to n\gamma$, $\Sigma^+\to p\gamma$, $\Sigma^0\to n\gamma$, $\Xi^0\to\Lambda\gamma$, $\Xi^0\to\Sigma^0\gamma$ and $\Xi^-\to\Sigma^-\gamma$. Using the notation defined in Fig.~\ref{Fig:HadronPlot}, these processes can be effectively described by the following Lagrangian:
 \begin{eqnarray}
{\cal L}=\frac{eG_F}{2}\bar{B}_f(k)(a+b\gamma_5)\sigma^{\mu\nu}B_i(p)F_{\mu\nu},\label{WRHDs:Lag}
\end{eqnarray}
where $a$ and $b$ with dimension of masses represent parity-conserving~($P$-wave) and parity-violating~($S$-wave) amplitudes, respectively. Here, $G_F$ is the Fermi constant, $e$ is elementary charge and $F_{\mu\nu}\equiv \partial_\mu A_\nu-\partial_\nu A_\mu$ where $A_\mu$ is the photon field. The terms $\sigma^{\mu\nu}$ and $\gamma_5$ represent combinations of  Dirac gamma matrices. The symbols $B_i$ and $B_f$ denote the octet baryons corresponding to the initial and final states, respectively.

Utilizing the Lagrangian given in Eq.~(\ref{WRHDs:Lag}), the differential decay rate of WRHDs that involve the polarization of the decaying baryons is typically expressed as 
\begin{eqnarray}
\frac{d\Gamma}{d\cos\theta}&=&\frac{e^2G_F^2}{\pi}(|a|^2+|b|^2)[1+\alpha_\gamma\cos\theta]\cdot|\vec{k}|^3,
\end{eqnarray}
with
\begin{eqnarray}
\alpha_\gamma\equiv\frac{2{\rm Re}(ab^*)}{|a|^2+|b|^2},\label{WRHDs:asypara}
\end{eqnarray}
where $\alpha_\gamma$ is the decay asymmetry parameter that describes the angular distribution of the decay relative to the
initial hyperon spin, and $\theta$ represents the angle between the initial hyperon spin and the final baryon 3-momentum $|\vec{k}|$.

WRHDs have remained a longstanding puzzle, both experimentally and theoretically. This puzzle is rooted in the so-called Hara theorem~\cite{Hara:1964zz}, which is based on three fundamental symmetry considerations: gauge invariance, conservation of the combined charge and parity symmetry ($CP$) and U-spin symmetry. These symmetries are vital for understanding the theoretical underpinnings of WRHDs. The first two of these symmetries are unambiguous and unshakable~\footnote{No signs of $CP$ violation in WRHDs have not been observed experimentally~\cite{BESIII:2022rgl,BESIII:2023fhs,BESIII:2024lio}.}. Furthermore, a discrete U-spin symmetry transformation, which interchanges $d$ and $s$ quarks, implies that the parity-violating decay amplitude $b$ must vanish for the $\Sigma^+\to p\gamma$ and the $\Xi^-\to\Sigma^-\gamma$ decays in the flavor SU(3) limit~\footnote{A detailed proof of this can be found in Refs.~\cite{Shi:2023kbu,Lach:1995we}}. When SU(3) symmetry breaking is accounted for, the expected range for $\alpha_\gamma(\Sigma^+\to p\gamma)$  is $\pm0.2$~\cite{Zenczykowski:2005cs}. However, the first experimental result in 1969 revealed an unexpectedly large and negative asymmetry, $\alpha_\gamma(\Sigma^+\to p\gamma)=-1.03_{-0.42}^{+0.52}$, which directly contradicts Hara's theorem. This unexpected result represents the initial puzzle in WRHDs. Following this, numerous phenomenological models~\cite{Gavela:1980bp,Nardulli:1987ub,Zenczykowski:1991mx,Zenczykowski:2005cs,Dubovik:2008zz,Niu:2020aoz,Xing:2023jnr} were proposed to investigate this discrepancy, some of which evaded Hara's theorem. While certain models could accommodate the large negative polarization asymmetries of $\Sigma^+\to p\gamma$, they failed to explain other WRHDs, such as $\Xi^0\to\Lambda\gamma$ and $\Xi^0\to\Sigma^0\gamma$~(see Refs.~\cite{Shi:2022dhw,Shi:2023kbu} and references therein). Notably, baryon chiral perturbation theory~($\chi$PT)~\cite{Jenkins:1992ab,Neufeld:1992np,Borasoy:1999nt} was also applied to study WRHDs but did not provide conclusive answers.

Recently, the BESIII Collaboration has reported new measurements of absolute branching fractions (BFs) and $\ar$s for the process  $\Sigma^+\to p\gamma$~\cite{BESIII:2023fhs}, $\Lambda\to n\gamma$~\cite{BESIII:2022rgl} and $\Xi^{0}\to\Lambda\gamma$ decays~\cite{BESIII:2024lio}. Especially, the $\ar$ for $\Lambda\to n\gamma$ was measured experimentally for the first time. These experimental results exacerbate the WRHD puzzle, as none of the existing theoretical frameworks can fully account for the observed data, although it is rather plausible that the Hara's theorem should not hold with the SU(3) flavor symmetry broken. The new measurements thus make it even more challenging to construct a unified theoretical model that can describe all the current observations of WRHDs compared to the initial puzzle's scope. In this review, we will briefly introduce the development of theoretical approaches, summarize the latest theoretical progress on WRHDs, assess the impact of recent BESIII measurements on our understanding of this puzzle, and discuss the experimental and theoretical challenges that lie ahead.

{\it{2.~Theory Development.}} Various theoretical approaches, both at the quark and hadron levels, have been employed over the years to investigate WRHDs. The quark-level methods provide specific dynamic mechanisms for addressing the WRHD puzzle and contribute to a deeper understanding of the internal structure of hadrons. However, numerous models-- such as the pole models (PM)~\cite{Gavela:1980bp,Nardulli:1987ub}, the vector-dominance model (VMD)~\cite{Zenczykowski:1991mx}, the broken SU(3) model (BSU(3))~\cite{Zenczykowski:2005cs}, the quark model~\cite{Dubovik:2008zz}, the non-relativistic constituent quark model~(NRCQM)~\cite{Niu:2020aoz}, and the light-front quark model~(LFQM)~\cite{Xing:2023jnr}--both with and without attempts to evade Hara's theorem, have all failed to adequately describe the WRHD data. An alternative approach is provided by model-independent methods at the hadron level, commonly referred to as effective field theories (EFTs). The key advantage of EFTs lies in their abilities to sidestep detailed assumptions about complex or unknown dynamic mechanisms. Through EFTs, physical quantities of interest can be computed with arbitrary precision using perturbative expansions, given that sufficient experimental data are available. Furthermore, EFTs can predict phenomena at higher energy scales or uncover evidence of new physics by operating within a defined low-energy regime. Among these,  $\chi$PT serves as the low-energy EFT of quantum chromodynamics (QCD) and has become a useful tool for studying WRHDs. Although $\chi$PT~\cite{Jenkins:1992ab,Neufeld:1992np,Borasoy:1999nt,Shi:2022dhw} has not resolved all experimental observations, recent developments~\cite{Shi:2022dhw} in covariant baryon chiral perturbation theory have brought it closer to addressing the WRHD puzzle compared to other models.

Moreover, WRHDs represent a valuable venue to search for $CP$ violation, a fundamental phenomenon that could explain the observed asymmetry between matter and antimatter in the universe~\cite{Sakharov:1967dj}. There has yet to have comprehensive studies on the $CP$ violation in WRHDs within the context of  $\chi$PT, largely due to the persistent S/P wave puzzle of hyperon non-leptonic decays~\cite{Jenkins:1991bt,Donoghue:1985rk,Holstein:2000yf}. In the bargain, only a few theoretical models predict $CP$ violation in WRHDs, with expected magnitudes at the order of $10^{-5}$ to $10^{-4}$ within the SM~\cite{Li:1993hp,Chen:1995vw,Xing:2023jnr}. It is noteworthy that experimental evidence for $CP$ violation in WRHDs has yet to be observed~\cite{BESIII:2023fhs,BESIII:2024lio}.

{\it{2.1~Application of NRCQM and LFQM to WRHD.}} Shortly before and after the release of the new BESIII data~\cite{BESIII:2022rgl,BESIII:2023fhs,BESIII:2024lio}, two theoretical studies revisited the WRHD puzzle within the frameworks of the  NRCQM~\cite{Niu:2020aoz} and the LFQM~\cite{Xing:2023jnr}. The NRCQM provides a natural mechanism for evading the Hara theorem and successfully describes the observed large negative asymmetry in $\Sigma^+\to p\gamma$. This study highlights the significance of pole term contributions to WRHDs, demonstrating that even a relatively small SU(3) flavor symmetry breaking can lead to substantial effects. These effects arise due to the interference between pole terms, thereby amplifying their contributions to the parity-violating amplitude of $\Sigma^+\to p\gamma$.

The predictions from the NRCQM for all WRHDs are summarized in \Cref{Tab: BR,Tab: Asymmetry}. From these results, it is evident that the BFs of most WRHDs align with  experimental data within their uncertainties, except for the $\Xi^0\to\Sigma^0\gamma$ decay. 
However, the $\ar$s for $\Lambda$ and $\Xi^0$ decays, as obtained by the NRCQM, do not agree with current experimental measurements. Specifically, even for the $\ar$s of $\Xi^0\to\Lambda\gamma$ and $\Xi^0\to\Sigma^0\gamma$, the NRCQM predicted the wrong signs. 

The newly developed LFQM method~\cite{Xing:2023jnr} adopts a three-quark picture, differing from  the diquark model employed by other LFQM theories~\cite{Wang:2022ias,Liu:2022mxv}. This approach enables the estimation of contributions from the $W$ boson exchange between two constituent quarks in the relevant Feynman diagrams. According to \Cref{Tab: BR,Tab: Asymmetry}, the only prediction from the LFQM that aligns with experimental data is $\alpha_\gamma(\Lambda\to n\gamma)$. Compared to the NRCQM results, the LFQM predicts larger BFs and smaller magnitudes of $\ar$s (in absolute value) for other WRHDs, except $\Xi^0\to\Sigma^0\gamma$. 

Although neither the NRCQM nor LFQM resolves the WRHD puzzle, the former underscores the significance of pole term contributions for achieving a unified theoretical description of all current WRHD observations. Their findings inspire further studies into exploring new dynamic mechanisms that might ultimately solve the WRHD puzzle. The key role played by the pole term contributions is also found in charmed baryon hadronic weak decays, e.g., $\Xi_c^0\to \Lambda_c\pi^-$~\cite{Niu:2021qcc}, and $\Lambda_c\to \Lambda\pi$ and $\Sigma\pi$~\cite{Niu:2020gjw}.

\begin{table}
  \centering
  \caption{ \label{Tab: BR} Experimental data and theoretical predictions of BFs ${\cal B}$ (in units of $10^{-3}$) of the WRHDs.}
  \begin{tabular}{cccc}
\hline
\hline
 ~Decay modes~ & ~Data~\cite{E761:1993unn,ParticleDataGroup:2024cfk,BESIII:2022rgl,BESIII:2023fhs,BESIII:2024lio}~ & ~NRCQM~\cite{Niu:2020aoz}~& ~LFQM~\cite{Xing:2023jnr}~\\
\hline
$\Lambda\to n\gamma$ & $0.832(38)(54)$ & $1.83(96)$ & $3.38$\\

$\Sigma^+\to p\gamma$ & $0.996(21)(18)$ & $1.06(59)$ & $2.81$\\

$\Sigma^0\to n\gamma$ & $\cdots$ & $10^{-10}$ & $2.79^{-10}$\\

$\Xi^0\to\Lambda\gamma$ & $1.347(66)(54)$ & $0.96(32)$ & $5.54$\\

$\Xi^0\to\Sigma^0\gamma$ & $3.33(10)$ & $9.75(418)$ & $8.12$\\

$\Xi^-\to \Sigma^-\gamma$ & $0.127(23)$ & $\cdots$ & $\cdots$\\
\hline
\hline
\end{tabular}
\end{table}
\begin{table*}[htb!]
  \centering
  \caption{ \label{Tab: Asymmetry} Experimental data and theoretical predictions of $\alpha_\gamma$s of the WRHDs.}
  \begin{tabular}{ccccc}
\hline
\hline
 ~~~Decay modes~~~ & ~~~Data~\cite{E761:1993unn,ParticleDataGroup:2024cfk,BESIII:2022rgl,BESIII:2023fhs,BESIII:2024lio}~~~ & ~~~NRCQM~\cite{Niu:2020aoz}~~~& ~~~LFQM~\cite{Xing:2023jnr}~~~ & ~~~EOMS $\chi$PT~\cite{Shi:2022dhw}~~~\\
\hline
$\Lambda\to n\gamma$ & $-0.16(10)(5)$ & $-0.67(6)$ & $-0.25$ & $[-0.43,~0.15]$\\

$\Sigma^+\to p\gamma$ & $-0.652(56)(20)$ & $-0.58(6)$ & $-0.1$ & $[-0.32,-0.27]$\\

$\Sigma^0\to n\gamma$ & $\cdots$ & $0.37(4)$ & $-0.22$ & $[-0.70,~0.70]$\\

$\Xi^0\to\Lambda\gamma$ & $-0.741(62)(19)$ & $0.72(11)$ & $0.23$ & $[-0.83,-0.59]$\\

$\Xi^0\to\Sigma^0\gamma$ & $-0.69(6)$ & $0.33(4)$ & $-0.15$ & $[-0.74,~-0.63]$\\

$\Xi^-\to \Sigma^-\gamma$ & $1.0(13)$ & $\cdots$ & $\cdots$ & $[-0.18,~0.38]$\\
\hline
\hline
\end{tabular}
\end{table*}

{\it{2.2~Application of EOMS $\chi$PT to WRHD.}} $\chi$PT is grounded in the spontaneous breaking of QCD chiral symmetry and provides a systematic framework for constructing effective Lagrangians at the hadron level and performing perturbative expansions in the low-energy regime. This approach circumvents the need for first-principles calculations that are computationally unfeasible, offering a useful tool for studying WRHDs.

Compared to earlier formulations of $\chi$PT~\cite{Jenkins:1992ab,Neufeld:1992np,Borasoy:1999nt}, a recent advancement~\cite{Shi:2022dhw} employs covariant baryon $\chi$PT with the extended-on-mass-shell~(EOMS) renormalization scheme. This updated study has revised the values of the relevant low energy constants, ensured consistent power counting, and determined counterterm contributions for the first time. In the EOMS $\chi$PT, hyperon non-leptonic decay results serve as key inputs for the analysis of WRHDs. Consequently, the EOMS $\chi$PT approach~\cite{Shi:2022dhw} reveals  correlations between BFs and $\ar$s due to the long-standing S/P wave puzzle in hyperon non-leptonic decays~\cite{Jenkins:1991bt,Donoghue:1985rk,Holstein:2000yf}. By constraining the experimental BFs, the EOMS $\chi$PT results for $\ar$s in each WRHD channel can be extracted, as summarized in \Cref{Tab: Asymmetry}. The findings demonstrate that EOMS $\chi$PT can simultaneously describe four weak radiative decay channels: $\Lambda\to n\gamma$, $\Xi^0\to\Lambda\gamma$, $\Xi^0\to\Sigma^0\gamma$ and $\Xi^-\to \Sigma^-\gamma$. However, the predictions exhibit a notable deviation from experimental results only for the $\Sigma^+\to p\gamma$ decay. Given the model-independent nature of $\chi$PT, future precision measurements of $\alpha_\gamma(\Xi^-\to \Sigma^-\gamma)$ could test the reliability of the current experimental results for $\Sigma^+\to p\gamma$, thereby contributing to a deeper understanding of the WRHD puzzle.

Furthermore, recent BESIII measurements~\cite{BESIII:2018cnd,BESIII:2021ypr,BESIII:2022qax} of the decay asymmetry parameters in non-leptonic decays of $\Lambda$ and $\Xi^-$ that are related to WRHDs exhibit significant deviations. Specifically, these deviations are at the order of $5\sigma$ for $\Lambda$ and approximately $2\sigma$ for $\Xi^-$. It is therefore of considerable interest to investigate whether similar deviations are observed in $\Sigma^+$ and $\Xi^0$ channels. Such deviations emphasize the need for careful treatment of non-leptonic hyperon decay measurements within the EOMS $\chi$PT. It is also worth noting that the latest analysis~\cite{Shi:2022dhw} does not account for contributions from intermediate resonances, such as the $\Lambda(1405)$ state. As discussed in prior studies~\cite{Borasoy:1999nt,Zenczykowski:1999vq}, intermediate resonances, which also corresponds to the pole terms in the NRCQM~\cite{Niu:2020aoz}, could play a significant role in WRHDs. Future studies with a self-consistent inclusion of such intermediate resonances in the same framework should be explored. 

{\it{3.~Experimental Development.}} Experimental studies of WRHDs began with bubble chamber
experiments~\cite{PhysRevLett.14.154,PhysRev.188.2077,Ang:1969hg,MANZ1980217} and have  continued using modern particle physics
spectrometers since then. These include both fixed target and collider
experiments~\cite{Bristol-Geneva-Heidelberg-Lausanne-QueenMaryColl-Rutherford:1985ksh,PhysRevLett.59.868,Hessey:1989ep,PhysRevLett.68.3004,PhysRevD.51.4638,BESIII:2023fhs,Noble:1992ya,Larson:1993ig,BESIII:2022rgl,PhysRevLett.64.843,NA48:1999dxg,NA48:2004gyg,Batley:2010bp,PhysRevLett.63.2717,PhysRevLett.86.3239,Bristol-Geneva-Heidelberg-Lausanne-QueenMaryColl-Rutherford:1986kqv,PhysRevLett.72.808,Bristol-Geneva-Heidelberg-Orsay-Rutherford-Strasbourg:1979bfc,BRISTOL-GENEVA-HEIDELBERG-ORSAY-RUTHERFORD-STRASBOURG:1984jku,E761:1994kzo}.
So far, the BFs and $\ar$s have been measured for the following WRHDs: $\Sigma^+\to
p\gamma$~\cite{Bristol-Geneva-Heidelberg-Lausanne-QueenMaryColl-Rutherford:1985ksh,PhysRevLett.59.868,Hessey:1989ep,PhysRevLett.68.3004,PhysRevD.51.4638,BESIII:2023fhs},
$\Lambda\to n\gamma$~\cite{Noble:1992ya,Larson:1993ig,BESIII:2022rgl}, $\Xi^{0}\to\Lambda\gamma$~\cite{PhysRevLett.64.843,NA48:1999dxg,NA48:2004gyg,Batley:2010bp},
$\Xi^0\to\Sigma^0\gamma$~\cite{NA48:1999dxg,Batley:2010bp,PhysRevLett.63.2717,PhysRevLett.86.3239}. For $\Xi^-\to
\Sigma^-\gamma$ only the BF has been measured~\cite{Bristol-Geneva-Heidelberg-Lausanne-QueenMaryColl-Rutherford:1986kqv,PhysRevLett.72.808}. 
The experimental results are summarized in \Cref{tab:hs1,tab:hs2,tab:hs3,tab:hs4,tab:hs5}. 
\begin{table}[htpb]
	\centering
	\caption{Summary of the experimental results of $\Sigma^+\to p\gamma$ decay}
	\label{tab:hs1}
        \begin{ruledtabular}
	\begin{tabular}{cccc}
		Date     & Experiment                             & BF ($\times 10^{-3}$)                        & $\alpha_\gamma$     \\
        \hline
		2023   & BESIII~\cite{BESIII:2023fhs}                               & \num{0.996 \pm 0.021}(18) & \num{-0.652 \pm
		0.056}(20)                                                                                                    \\
			1995   & E761~\cite{PhysRevD.51.4638}                                 & \num{1.20(0.08)}                             & -                   \\
		1992   & SPEC~\cite{PhysRevLett.68.3004}                                 & -                                            & \num{-0.720(0.086)} \\
		1989   & CNTR~\cite{Hessey:1989ep}                                 & \num{1.45(0.31)}                             & -                   \\
		1987   & CNTR~\cite{PhysRevLett.59.868}                                 & \num{1.23(0.20)}                             & -                   \\
		1985   & CNTR~\cite{Bristol-Geneva-Heidelberg-Lausanne-QueenMaryColl-Rutherford:1985ksh}                                 & \num{1.27(0.18)}                             & -                   \\
		1980   & HBC~\cite{MANZ1980217}                                  & \num{1.09(0.20)}                             & \num{-0.53(0.36)}   \\
		1969   & HBC~\cite{Ang:1969hg}                                  & \num{1.1(0.2)}                               & -                   \\
		1969   & HBC~\cite{PhysRev.188.2077}                                  & \num{1.42(0.26)}                             & \num{-1.03(0.52)}   \\
		1965   & HBC~\cite{PhysRevLett.14.154}                                  & \num{1.9(0.4)}                               & -                   \\
	\end{tabular}
                
        \end{ruledtabular}
\end{table}

\begin{table}[htpb]
	\centering
	\caption{Summary of the experimental results of $\Lambda\to n\gamma$ decay}
	\label{tab:hs2}
        \begin{ruledtabular}
	\begin{tabular}{cccc}
		Date     & Experiment                             & BF ($\times 10^{-3}$)                        & $\alpha_\gamma$     \\
        \hline
		2022 & BESIII~\cite{BESIII:2022rgl}   & \num{0.846 \pm 0.039}(52) & \num{-0.160 \pm 0.101}(46)                                                                                                               \\
			1994 & E761~\cite{Larson:1993ig}     & \num{1.75(0.15)}                                   & -                                             \\
			1992 & SPEC~\cite{Noble:1992ya}     & \num{1.78(0.24)}                                   & -                                             \\
	\end{tabular}
        \end{ruledtabular}
\end{table}

\begin{table}[htpb]
	\centering
	\caption{Summary of the experimental results of $\Xi^{0}\to \Lambda \gamma$ decay}
	\label{tab:hs3}
        \begin{ruledtabular}
	\begin{tabular}{cccc}
		Date     & Experiment                             & BF ($\times 10^{-3}$)                        & $\alpha_\gamma$     \\
        \hline
		2024 & BESIII~\cite{BESIII:2024lio}     & \num{1.347\pm0.066}(54)                     & \num{-0.741\pm0.062}(19) \\
		2010 & NA48~\cite{Batley:2010bp}     & -                     & \num{-0.704(0.064)} \\
		2004 & NA48~\cite{NA48:2004gyg}     & \num{1.17(0.09)}      & \num{-0.78(0.18)}   \\
		2000 & NA48~\cite{NA48:1999dxg}     & \num{1.91(0.34)}      & -                   \\
		1990 & SPEC~\cite{PhysRevLett.64.843}     & \num{1.06(0.18)}      & \num{-0.43(0.44)}   \\
	\end{tabular}
                
        \end{ruledtabular}
\end{table}

\begin{table}[htpb]
	\centering
	\caption{Summary of the experimental results of $\Xi^0\to \Sigma^0\gamma$ decay}
	\label{tab:hs4}
    \begin{ruledtabular}

	\begin{tabular}{cccc}
		Date     & Experiment                             & BF ($\times 10^{-3}$)                        & $\alpha_\gamma$     \\
		\hline
		2010 & NA48~\cite{Batley:2010bp}     & -                     & \num{-0.729(0.076)} \\
		2001 & KTEV~\cite{PhysRevLett.86.3239}     & \num{3.34(0.09)}      & \num{-0.63(0.09)}   \\
		2000 & NA48~\cite{NA48:1999dxg}     & \num{3.16(0.76)}      & -                   \\
		1989 & SPEC~\cite{PhysRevLett.63.2717}     & \num{3.56(0.42)}      & \num{0.20(0.32)}    \\
	\end{tabular}
            
    \end{ruledtabular}
\end{table}

\begin{table}[htpb]
	\centering
	\caption{Summary of the experimental results of $\Xi^{-}\to\Sigma^{-}\gamma$ decay}
	\label{tab:hs5}
    \begin{ruledtabular}

	\begin{tabular}{cccc}
		Date     & Experiment                             & BF ($\times 10^{-3}$)                        & $\alpha_\gamma$     \\
		\hline
		1994 & E761~\cite{PhysRevLett.72.808}     & \num{0.122(0.023)}    & -               \\
		1987 & SPEC~\cite{Bristol-Geneva-Heidelberg-Lausanne-QueenMaryColl-Rutherford:1986kqv}     & \num{0.227(0.102)}    & -               \\
	\end{tabular}
            
    \end{ruledtabular}
\end{table}

\begin{figure}[htpb]
	\centering
	\includegraphics[width=0.3\linewidth]{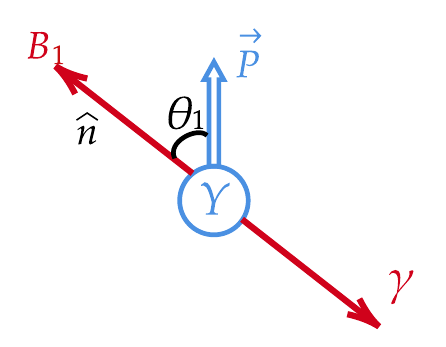}
        \put(-60,0){(a)}
        \includegraphics[width=0.3\linewidth]{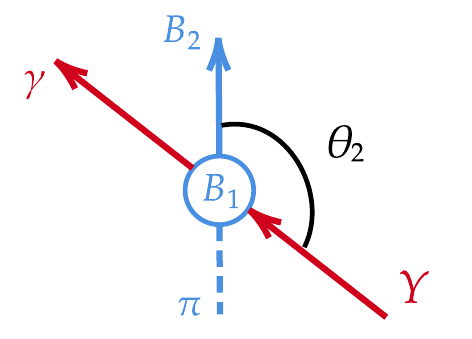}
        \put(-60,0){(b)}
	\caption{(a) Schematic diagram of a polarized hyperon ($Y$) decaying into a stable baryon ($B_1$) and a photon. The polarization vector of the hyperon
	($\vec{P}$), the direction vector of the daughter baryon ($\hat{n}$) and the opening angle ($\theta_1$) are also represented. (b) Schematic diagram of the cascade decay $Y\to B_1 \gamma, B_1\to B_2 \pi$ in the rest frame of $B_1$. }%
	\label{fig:diagram1}
\end{figure}

For experiments other than electron-positron
collision experiments, hyperons are typically produced through meson-nucleon scatterings (e.g. $K^- p\to \pi^- \Sigma^+$) or through interactions between a hadron beam and a target. The production momentum of hyperons in these processes can vary
from zero to hundreds of \si{GeV/\clight}. Due to the difficulty in precisely measuring the production rate within the detector acceptance, the
BF of a WRHD is usually normalized to the BF of a reference decay, which is generally a dominant decay mode of the same hyperon. This is commonly referred to as the
relative BF measurement. The $\alpha_\gamma$s of  WRHDs can be measured in two distinct ways, depending on the specific decay modes. If the daughter baryon is stable (e.g., the proton in
$\Sigma^+\to p\gamma$ decay), the mother hyperon must have transverse polarization for the $\alpha_\gamma$ measurement. In this scenario, the angular distribution of the daughter baryon in the rest frame of the decaying hyperon is given by:
\begin{equation}
	\frac{\dif N}{\dif \Omega} = \frac{N_0}{4\pi}(1+\alpha_\gamma \vec{P}\cdot\hat{n}) =   \frac{N_0}{4\pi}(1+\alpha_\gamma f_{\mathrm{pol}}\cos\theta_1),
\end{equation}
where $\vec{P}$ is the transverse polarization vector of the hyperon, $\hat{n}$ is the direction vector of the daughter baryon in the rest frame of the
hyperon, $\theta_1$
is the angle between $\vec{P}$ and $\hat{n}$, and $f_{\mathrm{pol}}$ is the magnitude of the hyperon's transverse polarization. The variables involved are illustrated in \Cref{fig:diagram1}~(a).  The transverse polarization of the hyperon originates from the production process and can typically only be measured with a precision
of $\mathcal{O}(10)\%$. Therefore, experiments often focus on measuring the relative value of $\alpha_\gamma$ with respect to a reference decay of the same
hyperon. This approach mitigates uncertainties arising from factors such as hyperon polarization, detection efficiency asymmetry, and other systematic effects. As a result, the major source of systematic uncertainty in these experiments stems from the uncertainties associated with the reference decay asymmetry parameter.
When the daughter baryon is unstable and possesses its own decay asymmetry parameter $\alpha_0$, the $\alpha_\gamma$ of the mother hyperon can be measured
independently of transverse polarization. Specifically, \Cref{fig:diagram1}~(b) illustrates the schematic diagram of a cascade decay process $Y\to B_1 \gamma, B_1\to B_2 \pi$ in the rest
frame of $B_1$. The angular distribution of $B_2$ in this reference frame can be expressed as:
\begin{equation}
	\frac{\dif N}{\dif \cos\theta_2} = N_0(1-\alpha_\gamma \alpha_0\cos\theta_2),
\end{equation}
where $\theta_2$ is defined as the opening angle between the mother hyperon and daughter $B_2$ in the rest frame of $B_1$. This allows for the extraction of the product of $\alpha_\gamma$ and $\alpha_0$.
However,  measuring the absolute value of BF and $\alpha_\gamma$ in such experiments remains a significant challenge.

Electron-positron collision experiments offer a unique and powerful method for investigating these WRHDs.
The BESIII detector~\cite{BESIII:2009fln} records symmetric $e^+e^-$ collisions
provided by the BEPCII storage ring~\cite{Yu:IPAC2016-TUYA01}
in the center-of-mass energy range from \num{1.84} to \SI{4.95}{GeV},
with a peak luminosity of $1.1 \times 10^{33}\;\text{cm}^{-2}\text{s}^{-1}$
achieved at $\sqrt{s} = 3.773\;\text{GeV}$.
The BESIII experiment has amassed extensive datasets in this energy range~\cite{Ablikim_2020}. 
At \SI{3.097}{GeV}, corresponding to the peak production rate of the $\jpsi$ resonance, BESIII has recorded the world’s largest
$\jpsi$ data sample, consisting of \num{1.0087\pm0.0044e10} events~\cite{BESIII:2021cxx}. Additionally, a total of \num{2712.4\pm14.3 e6} $\psi(2S)$ events
have been collected around \SI{3.686}{GeV}~\cite{BESIII:2024lks}. These collision experiments are particularly advantageous for studying WRHDs because hyperons are produced in pairs through $\jpsi$ and $\psi(2S)$ decays, providing unique opportunities for the detailed studies.

\begin{figure}[htpb]
	\centering
	\includegraphics[width=\linewidth]{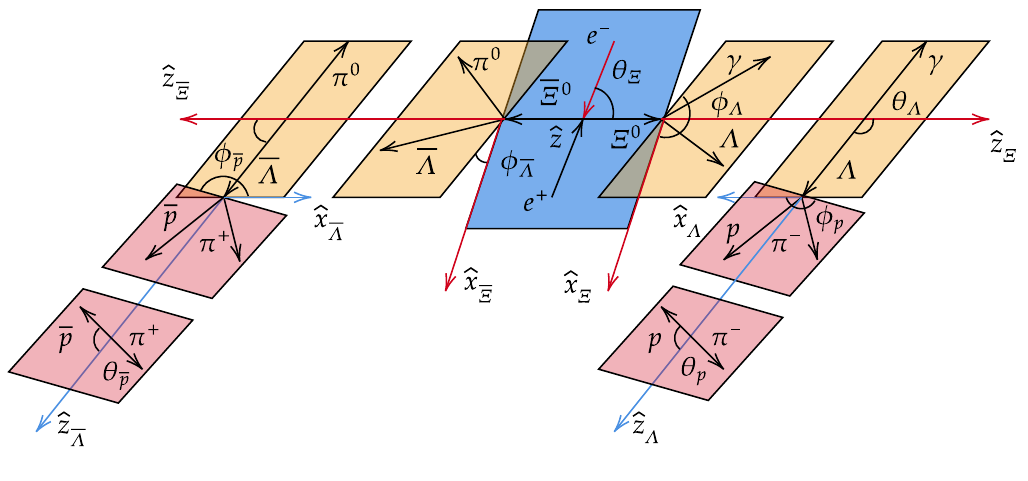}
	\caption{The helicity angles and helicity frames in the decay $\eexi$. The decay planes of $\xiz$ ($\xib$) and $\Lambda$ ($\bar{\Lambda}$) in the lab
		frame and in their mother particles' helicity frame are simultaneously shown in the figure, colored in orange and pink, respectively.}%
	\label{fig:frame3}
\end{figure}

As an illustrative example, the decay $\XiRadDec$ showcases how BESIII leverages these opportunities. Spin-correlated $\Xi^0\bar{\Xi}^0$ pairs are generated from the decay of $\jpsi$. The
$\bar{\Xi}^0$s decaying into $\bar{\Lambda}\piz$, $\bar{\Lambda}\to\bar{p}\pi^+$ are selected as the tag mode for $\Xi^0\bar{\Xi}^0$ events, and the $\Xi^0$s
decay via the process $\XiRadDec$. The full decay chain is presented in \Cref{fig:frame3}. 

In this case, the absolute BF can be measured using the double tag (DT) method. In this approach, the four momentum
of $\bar{\Lambda}$ and $\piz$ is reconstructed, and $\bar{\Xi}^0$ is 
reconstructed through its invariant mass: $M_{\bar{\Lambda}\piz} = \sqrt{(E_{\bar{p}\pi^+}+E_{\piz})^2  -(\mathbf{p}_{\bar{p}\pi^+}+\mathbf{p}_{\piz})^2
			      }$,
where $E_{\bar{p}\pi^+}$ and $E_{\piz}$ are the energy of $\bar\Lambda$ and $\piz$ candidates, respectively, and $\mathbf{p}_{\bar{p}\pi^+}$ and $\mathbf{p}_{\piz}$ are their corresponding momentum vectors. The events $\jpsi\to\Xi^0\bar{\Xi}^0$ are then identified by analyzing the recoil mass of $\bar{\Xi}^0$, defined as: $\mrec =\sqrt{(E_{\mathrm{cm}}-E_{\bar{p}\pi^+}-E_{\piz})^2  -(\mathbf{p}_{\bar{p}\pi^+}+\mathbf{p}_{\piz})^2 }$,
where $E_{\mathrm{cm}}$ is the CME. Events selected through this method are referred to as single tag (ST) events, and their number ($N_{\mathrm{ST}}$)
can be expressed as: $N_{\mathrm{ST}} = N_{\jpsiXX}\times  \mathcal{B}_{\XibHadDec} \times \mathcal{B}_{\LbDec}\times \varepsilon_{\mathrm{ST}}$, where $N_{\jpsiXX}$ is the number of $\jpsiXX$ events in the dataset, $\mathrm{BF}_{\XibHadDec}$ ($\mathrm{BF}_{\LbDec}$) is the BF of
$\XibHadDec$ ($\LbDec$), and  $ \varepsilon_{\mathrm{ST}}$ is the detection efficiency for the  ST events. Once $N_{\mathrm{ST}}$ is established, the number of recoiled $\Xi^0$ candidates can be counted in a model-independent manner. The signal of interest corresponding to $\XiRadDec$  is then searched for by reconstructing events in which $\Xi^0$ recoils against the ST $\bar{\Xi}^0$. This results in a DT sample, and the number of DT events can be expressed as: $N_{\mathrm{DT}} = N_{\jpsiXX}\times \mathrm{BF}_{\XibHadDec}\times\mathrm{BF}_{\LbDec} \times \mathrm{BF}_{\XiRadDec}\times\mathrm{BF}_{\LDec} \times
	\varepsilon_{\mathrm{DT}}$,
where $\mathrm{BF}_{\XiRadDec}$ ($\mathrm{BF}_{\LDec}$) is the BF of  $\XiRadDec$  ($\LDec$), and $\varepsilon_{\mathrm{DT}}$ is the DT detection
efficiency.  
Using the DT method, the BF of the decay $\XiRadDec$ can be computed with:
\begin{equation}
	\mathrm{BF}_{\XiRadDec}= \frac{N_{\mathrm{DT}}} {N_{\mathrm{ST}}} \times \frac{\varepsilon_{\mathrm{ST}}} {\mathrm{BF}_{\LDec}\varepsilon_{\mathrm{DT}}}.
\end{equation}
The DT method is particularly advantageous because it cancels systematic uncertainties arising from ST selection. Consequently, the measured results are independent of the BF
measurement results of $\XibHadDec$ and $\jpsiXX$, achieving an absolute measurement of the BFs of interest.

The $\alpha_\gamma$ of WRHDs can be extracted from the full decay chain's amplitude~\cite{FALDT201716,PhysRevD.99.056008,Batozskaya:2023rek}. Using the helicity formalism of Jacob and Wick~\cite{Jacob:1959at}, the general
expression for the decay amplitude of $\eexi$ can be written as:
\begin{equation}
	\label{eq:a}
	\mathcal{W} = \sum^{3}_{\mu,\bar{\nu}=0}\sum^{3}_{\mu^\prime, \bar{\nu}^\prime=0}
	C_{\mu\bar{\nu}}a_{\mu\mu^\prime}^{\Xi}a_{\mu^\prime0}^{\Lambda}a_{\bar{\nu}\bar{\nu}^\prime}^{\bar{\Xi}}a_{\bar{\nu}^\prime0}^{\bar{\Lambda}}.
\end{equation}
In the formula, $C_{\mu\nu}$ represents the polarization and spin correlation of $\xiz$ and $\xib$, $a_{\mu\mu\prime}$ represents the decay matrix of
hyperons
($i.e.$ $\xiz$, $\xib$, $\Lambda$ and $\bar{\Lambda}$) depending on the decay products of the hyperon. The explicit expressions for
$C_{\mu\nu}$ are given below:
\begin{equation}
	\begin{aligned}
		\label{eq:cuv}
		C_{00} & =2(1+\alpha_{\psi}\cos^2\theta_{\xiz}),                                              \\
		C_{02} & =2\sqrt{1-\alpha^2_{\psi}}\sin\theta_{\xiz}\cos\theta_{\xiz}\sin(\Delta\Phi_{\psi}), \\
		C_{11} & =2\sin^2\theta_{\xiz},                                                               \\
		C_{13} & =2\sqrt{1-\alpha^2_{\psi}}\sin\theta_{\xiz}\cos\theta_{\xiz}\cos(\Delta\Phi_{\psi}), \\
		C_{20} & =-C_{02},                                                                            \\
		C_{22} & =\alpha_{\psi}C_{11},                                                                \\
		C_{31} & =-C_{13},                                                                            \\
		C_{33} & =-2(\alpha_{\psi}+\cos^2\theta_{\xiz}),                                              \\
	\end{aligned}
\end{equation}
where $\alpha_{\psi}$ and $\Delta\Phi_{\psi}$ are the decay parameters of the decay $\jpsiXX$. The decay matrix of hyperon  $a_{\mu\mu\prime}$  depends on the decay products due to the different spin and parity of $\pi^{\pm,0}$ and photon. The $\ar$ of interested is involved in the radiative decay matrix as:
\begin{equation}
	a_{\gamma}=\left(\begin{matrix}
			1                        & 0 & 0 & -\alpha             \\
			\alpha\cos\phi\sin\theta & 0 & 0 & -\sin\theta\cos\phi \\
			\alpha\sin\theta\sin\phi & 0 & 0 & -\sin\theta\sin\phi \\
			\alpha\cos\theta         & 0 & 0 & -\cos\theta         \\
		\end{matrix}
	\right),
\end{equation}
where $\theta$ and $\phi$ are the helicity angles of the daughter baryon. The
definitions
of helicity angles are illustrated in \Cref{fig:frame3}.

The decay parameters $\alpha_\psi$ and $\Delta\Phi_\psi$ are directly related to the psionic electromagnetic form factors $G_E^\psi$, $G_M^{\psi}$ of the process $\ee\to\jpsi\to\Xi^0\bar{\Xi}^0$~\cite{FALDT201716}. Specifically, $\alpha_\psi$ represents the magnitude difference between $G_E^\psi$ and $G_M^{\psi}$ related amplitudes, and $\Delta\Phi_\psi$ denotes the relative phase between these two form factors. According to \Cref{eq:cuv}, a nonzero $\Delta\Phi_\psi$ introduces transverse polarization to the produced hyperons, which is proportional to
$\frac{\sqrt{1-\alpha_\psi^2}\cos\theta_{\Xi^{0}}\sin\theta_{\Xi^{0}}}{1+\alpha_\psi\cos^2\theta_{\Xi^{0}}}\sin\Delta\Phi_\psi$. Furthermore, by incorporating the full decay chain's amplitude, the spin correlations between the paired hyperons are established. These spin correlations impose stringent constraints on the helicity angle distributions, thereby significantly enhancing the sensitivity of $\alpha_\gamma$ measurements.  Thanks to the effects of transverse polarization and spin
correlation of the hyperons, the precision of $\alpha_\gamma$ measurements in WRHDs can be substantially improved by several times under the same statistical conditions, compared to fixed-target experiments\cite{BESIII:2023fhs,Hong:2023soc}. 

\begin{figure}[htpb]
	\centering
	\includegraphics[width=\linewidth]{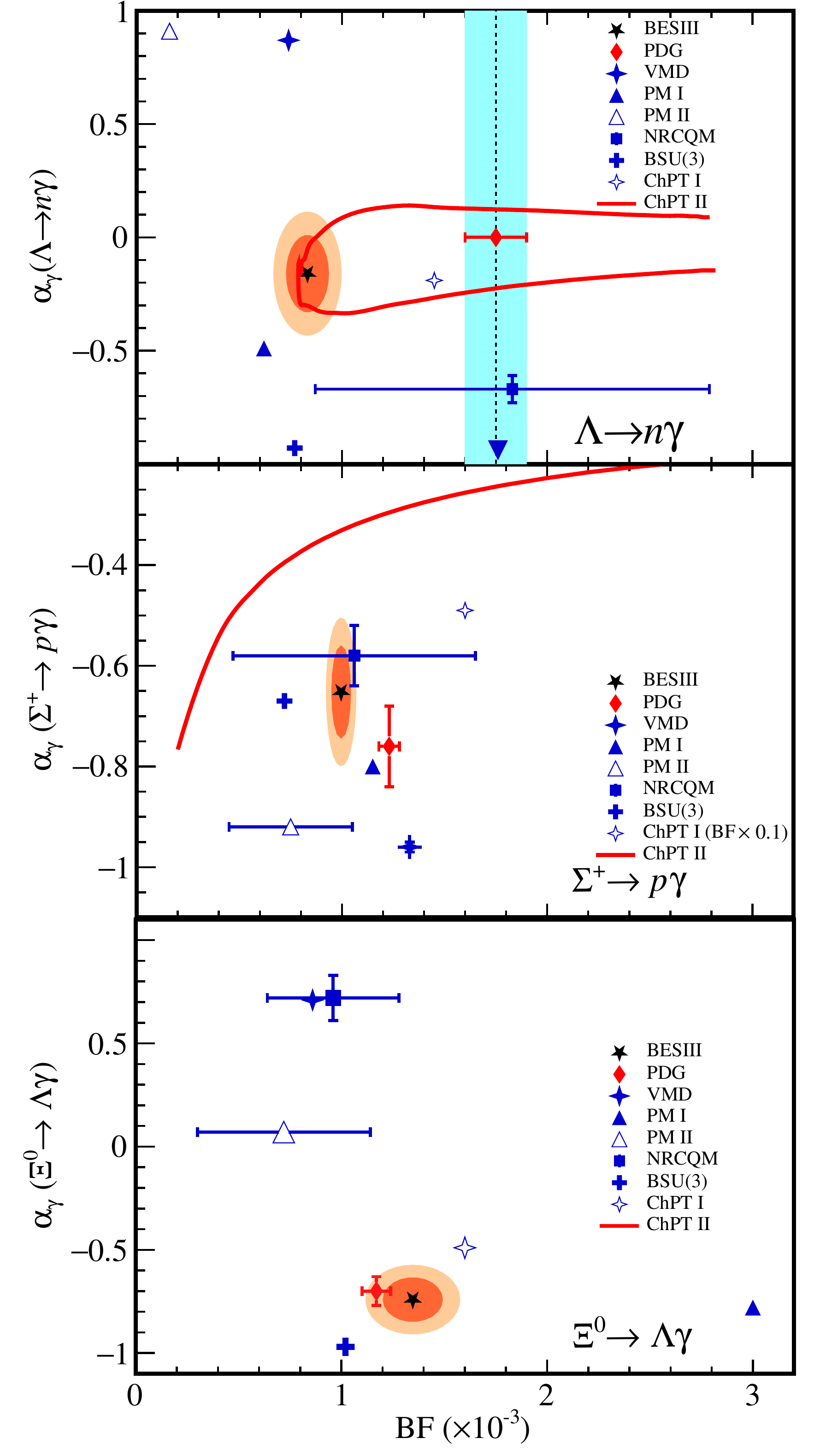}
	\caption{Distribution of $\alpha_{\gamma}$ versus BF of the $\Lambda\to n\gamma$, $\SigRadDec$ and $\XiRadDec$ decays. The black marks denote the results measured by	BESIII and the orange contours correspond to the	$\SI{68}{\percent}/\SI{95}{\percent}$ confidence-level of the	results. The red diamonds represent the PDG	values~\cite{ParticleDataGroup:2024cfk} of the BF and $\alpha_\gamma$. The red colored lines show the predicted $\alpha_{\gamma}$ as a function of BF cited from Ref.~\cite{Shi:2022dhw}. Other symbols stand for the results predicted by the VMD~\cite{zenczykowski_reanalysis_1991}, PM (PM I refers	to Ref.~\cite{gavela_parity_1981} and PM II refers to	Ref.~\cite{nardulli_pole_1987}), NRCQM~\cite{Niu_2020}, BSU(3)~\cite{zenczykowski_joint_2006} and another $\chi$PT model~\cite{borasoy_resonances_1999}.}%
	\label{fig:com}
\end{figure}

The notable advancements in BF and $\alpha_\gamma$ measurement derived from $\jpsi$ decays underscore the strong motivation for experimental investigations in this area. To date, the BESIII collaboration has achieved world-leading precision in measuring the BFs and
$\alpha_\gamma$s of $\Lambda\to n\gamma$~\cite{BESIII:2022rgl}, $\Sigma^+\to p\gamma$~\cite{BESIII:2023fhs} and $\XiRadDec$~\cite{BESIII:2024lio}. 
The measured values of these BFs and $\alpha_\gamma$s are summarized in \Cref{tab:hs1,tab:hs2,tab:hs3}. A comparison of the measured results with the world average values~\cite{ParticleDataGroup:2024cfk} and theoretical predictions is presented in \Cref{fig:com}. Additionally, the diviations between the measured results from BESIII and the world averages are systematically summarized in \Cref{tab:div}.

\begin{table}[htpb]
	\centering
	\caption{Divergence between the BESIII results and the world average values}
	\label{tab:div}
    \begin{ruledtabular}

	\begin{tabular}{ccc}
		Process & BF divergence ($\times\sigma$) & $\alpha_\gamma$ divergence ($\times\sigma$) \\
		\hline
		$\Lambda\to n\gamma$ & 5.6 & --- \\
		$\Sigma^+\to p\gamma$ & 4.2 & 1.1 \\
		$\XiRadDec$ & 1.6& 0.6\\
	\end{tabular}
            
    \end{ruledtabular}
\end{table}

The pair production of hyperons and anti-hyperons at the BESIII experiment enables the simultaneous measurement of charge-conjugate BF and $\alpha_\gamma$ within the same data
sample. This approach offers a significant reduction in systematic uncertainties when comparing results across charge-conjugated decay channels, a feature that is especially important for exploring $CP$ violation.
While weak radiative decays of heavy mesons have been extensively studied~\cite{E731:1992ace,BaBar:2012fqh,Belle:2014nmp,Belle:2016mtj,Belle:2017hum}, measurements on the $CP$ violation of such processes in the baryon sector are still very limited~\cite{LHCb:2021byf,BESIII:2023fhs}. WRHDs  are  experimentally compelling due to their large BFs at the level of $\mathcal{O}(10^{-3})$ and high production rate, compared to the weak radiative decays of other heavy flavor baryons.
The BESIII experiment has investigated the $CP$ asymmetry of WRHDs in the decays $\SigRadDec$ and $\XiRadDec$ by defining two $CP$ observables:
\begin{equation}
	\begin{aligned}
		 & \Delta_{CP}=\frac{\mathrm{BF}_+-\mathrm{BF}_-}{\mathrm{BF}_++\mathrm{BF}_-},       \\
		 & A_{CP}=\frac{\alpha_\gamma+\bar{\alpha}_\gamma}{\alpha_\gamma-\bar{\alpha}_\gamma}, \\
	\end{aligned}
\end{equation}
where $\mathrm{BF}_+$ ($\mathrm{BF}_-$) denotes the BF of the WRHD of hyperon (anti-hyperon), and $\bar{\alpha}_\gamma$ is the decay asymmetry parameter of the
WRHD of the anti-hyperon. The measured $CP$ asymmetry of $\SigRadDec$ and $\XiRadDec$ decays are summarized in \Cref{tab:cp}. This is the first time that the $CP$
asymmetry of WRHDs has been measured, and no excess $CP$ asymmetry beyond the SM prediction has been observed. The BF of the $\XiRadDec$ decay remains invariant under the combined $CP$ transformation;
therefore, it's not considered in this review~\cite{Bigi:2021hxw}.
\begin{table}[htpb]
	\centering
	\caption{$CP$ asymmetry of WRHDs measured by BESIII. The first uncertainty is statistical and the second is systematic.}
	\label{tab:cp}
    \begin{ruledtabular}

	\begin{tabular}{ccc}
		Process & $\Delta_{CP}$ & $A_{CP}$ \\
        \hline
		$\SigRadDec$ & $0.006(11)(4)$ &	$0.095(49)(31)$ \\
		$\XiRadDec$ & -- & $-0.120(84)(29)$ \\
	\end{tabular}
            
    \end{ruledtabular}
\end{table}

{\it{4.~Summary and Prospect.}}
As summarized in this review, weak hyperon decays serve as an ideal framework for exploring the non-perturbative aspects of strong interactions and probing potential new physics beyond the SM. In particular, the study of WRHDs remains challenging, with numerous theoretical approaches, such as $\chi$PT and other phenomenological frameworks, striving to elucidate their underlying mechanisms. Despite these efforts, current theoretical predictions have yet to fully reconcile with available experimental data. Notably, a recent study based on the EOMS $\chi$PT theory~\cite{Shi:2022dhw} offers a more promising resolution to the WRHD puzzle than alternative models. Additionally, $CP$ violation in WRHDs has been briefly reviewed, highlighting the fact that current experimental measurements remain one to two orders of magnitude short of the SM predictions.

From a theoretical perspective, two promising avenues could contribute to resolving the WRHD puzzle. One is the incorporation of novel dynamical mechanisms into existing models, while the other involves explicitly accounting for intermediate resonance contributions within the EOMS $\chi$PT framework.

Future theoretical advancements must be accompanied by corresponding experimental progress. Ongoing and forthcoming studies on WRHDs and nonleptonic hyperon decays are expected to provide crucial insights into the nature of these processes. One particularly significant measurement is $\alpha_\gamma(\Xi^-\to \Sigma^-\gamma)$, which could serve as a critical test of Hara’s theorem and further clarify the theoretical challenges associated with WRHDs. 
Although not discussed in detail in this review, further studies of weak radiative decays of spin-3/2 hyperons will be essential to complement ongoing investigations. Regretfully, these processes suffer from significantly lower branching fractions, posing considerable experimental challenges while potentially offering greater sensitivity to new physics.

A major advancement in addressing these experimental challenges would be the proposed super tau-charm facility~\cite{Charm-TauFactory:2013cnj,Achasov:2023gey}, which is expected to increase the available dataset by a factor of 100 compared to BESIII. This substantial enhancement in hyperon decay statistics from charmonium decays will significantly reduce statistical uncertainties for all measured BFs, $\alpha_\gamma$s, and $CP$ asymmetries in WRHDs, potentially by one order of magnitude. The anticipated results from this facility will be pivotal in advancing our understanding of WRHDs, with more precise measurements offering valuable insights into potential $CP$ violation in these decays.

{\it{Acknowledgments.}}
The authors (Z.K. Jia, X.R. Zhou, H.P. Peng) thank the supercomputing center of
USTC for their strong support. This work is supported by National Key R\&D Program
of China under Contracts Nos. 2023YFA1609400, 2023YFA1606703, 2020YFA0406400, 2020YFA0406300, ; 
International partnership program of the Chinese Academy of Sciences Grant
No. 211134KYSB20200057;
National Natural Science Foundation of China (NSFC) under Contracts Nos. 12035013, 12061131003, 12122509, 12105276, 11625523, 12405091, 12235018, 12435007; CAS Youth Team Program under Contract No. YSBR-101.

\bibliography{WRHDsCPL}

\end{document}